\documentclass[aps,prd,twocolumn,showpacs,superscriptaddress]{revtex4}

\usepackage{graphicx}
\usepackage{epsfig}
\usepackage{subfigure}

\usepackage{amsmath,amsthm,amsfonts,latexsym}

\newcommand{\ket}[1]{| #1 \rangle}

\newcommand{\matx}[1]{| #1 \rangle \langle #1 |}

\begin{document}

%\preprint{}

%\title{Global asymmetry of $SL(2,\mathbb{C})$ invariant two-qubit correlations from a Lorentz group gauge theory}
%\title{Quantum correlations can be twisted: A lattice gauge theory approach to SLOCC invariants of many qubits}% \emph{or}
\title{Global asymmetry of many-qubit
correlations: A lattice gauge theory approach}
%Local $SL(2,\mathbb{C})$ invariants for multi-qubit states from a Lorentz group gauge theory}


%\author{Mark S. Williamson$^{(1,2)}$, Marie Ericsson$^{(3)}$, Markus Johansson$^{(3)}$, Erik Sj\"{o}qvist$^{(1,3)}$, Anthony Sudbery$^{(4)}$, Vlatko Vedral$^{(1,5)}$ and William K. Wootters$^{(6)}$}
%\thanks{}
%\affiliation{$^{(1)}$Centre for Quantum Technologies, National University of Singapore, 3 Science Drive 2, Singapore 117543,\\
%$^{(2)}$Erwin Schr\"{o}dinger International Institute for Mathematical Physics, Boltzmanngasse 9, 1090 Wien, Austria,\\
%$^{(3)}$Department of Quantum Chemistry, Uppsala University, Box 518, SE-751 20 Uppsala, Sweden,\\
%$^{(4)}$Department of Mathematics, University of York, Heslington, York YO10 5DD, UK,\\
%$^{(5)}$Clarendon Laboratory, University of Oxford, Parks Road, Oxford OX1 3PU, UK,\\
%$^{(6)}$Department of Physics, Williams College, Williamstown, Massachusetts 01267, USA.
%}

\author{Mark S. Williamson}\email{m.s.williamson04@gmail.com}
\affiliation{Erwin Schr\"{o}dinger International Institute for Mathematical Physics, Boltzmanngasse 9, 1090 Wien, Austria,}
\affiliation{Centre for Quantum Technologies, National University of Singapore, 3 Science Drive 2, Singapore 117543,}
\author{Marie Ericsson}
\affiliation{Department of Quantum Chemistry, Uppsala University, Box 518, SE-751 20 Uppsala, Sweden,}
\author{Markus Johansson}
\affiliation{Department of Quantum Chemistry, Uppsala University, Box 518, SE-751 20 Uppsala, Sweden,}
\author{Erik Sj\"{o}qvist}
\affiliation{Centre for Quantum Technologies, National University of Singapore, 3 Science Drive 2, Singapore 117543,}
\affiliation{Department of Quantum Chemistry, Uppsala University, Box 518, SE-751 20 Uppsala, Sweden,}
\author{Anthony Sudbery}
\affiliation{Department of Mathematics, University of York, Heslington, York YO10 5DD, UK,}
\author{Vlatko Vedral}
\affiliation{Centre for Quantum Technologies, National University of Singapore, 3 Science Drive 2, Singapore 117543,}
\affiliation{Clarendon Laboratory, University of Oxford, Parks Road, Oxford OX1 3PU, UK.}
%\author{William K. Wootters}
%\affiliation{Department of Physics, Williams College, Williamstown, Massachusetts 01267, USA.}

\date{\today}

\begin{abstract}
We introduce a novel bridge between the familiar gauge field theory approaches used in many areas of modern physics such as quantum field theory and the SLOCC protocols familiar in quantum information. Although the mathematical methods are the same the meaning of the gauge group will be different. The measure we introduce, `twist', is constructed as a Wilson loop from a correlation induced holonomy. The measure can be understood as the global asymmetry of the bipartite correlations in a loop of three or more qubits; if the holonomy is trivial (the identity matrix), the bipartite correlations can be globally untwisted using general local qubit operations, the gauge group of our theory, which turns out to be the group of Lorentz transformations familiar from special relativity. If it is not possible to globally untwist the bipartite correlations in a state globally using local operations, the twistedness is given by a non-trivial element of the Lorentz group, the correlation induced holonomy. We provide several analytical examples of twisted and untwisted states for three qubits, the most elementary non-trivial loop one can imagine.
\end{abstract}
\maketitle

\section{Introduction}

There has been a large effort during the last decade to study the properties of entanglement and correlations in quantum states. Progress has been made quantifying and characterizing these correlations for composite states of two and more qubits \cite{ref:Horodecki07}. However, for states of three or more qubits the problem becomes hard as the possible ways a state may be entangled becomes greater. In this paper we introduce a new approach in the spirit of a lattice gauge field theory. Our motivation to use a gauge theory approach comes from several lines: (i) Our most successful theories of nature are expressed in this language and are well developed mathematically. Some examples of gauge theories in physics include classical electromagnetism, quantum field theories (such as quantum electrodynamics and quantum chromodynamics), string theory and general relativity. Mapping the study of quantum correlations to a gauge theory may allow one to use some of the techniques previously developed. (ii) Gauge theories have a natural geometric interpretation with which to gain intuition about a problem, the gauge field defining a curved surface and the Wilson loop, a gauge invariant observable of the theory, giving a measure of the total curvature of the gauge field. In our work, the gauge group that emerges naturally is the Lorentz group.

The usual scenario one considers when studying entanglement is the following: There are $N$ parties in spatially separated locations each holding one part of a composite quantum state comprised of $N$ subsystems. Each party is free to make local operations on their part of the state and classically communicate this operation and its outcome to the other parties in an effort to put the state in a standard form. They can use deterministic operations given by unitary groups which are simple rotations of their subsystems or more general local operations given by Kraus operators that may only be successful probabilistically. If two different states can be brought to the same standard form reversibly they are said to belong to the same entanglement class. If the conversion is deterministic, the two states are unitarily equivalent. If the conversion is only probabilistic, the two states are said to be equivalent under stochastic local operations and classical communication (SLOCC) \cite{ref:Bennett00}. In this paper we concentrate on the latter scenario for qubits. In this case the group of reversible local qubit operations is $SL(2,\mathbb{C})$ up to a positive constant less than or equal to unity. Moreover, it is known that the entanglement measures concurrence \cite{ref:Wootters98} and three-tangle \cite{ref:Coffman00} are invariant under the action of $SL(2,\mathbb{C})$.

There are a number of results concerning SLOCC classification. D\"{u}r \emph{et al.} \cite{ref:Dur00} have shown that there exist two classes for pure states of three qubits, the Greenberger-Horne-Zeilinger (GHZ) class and the W class. Verstraete \emph{et al.} \cite{ref:Verstraete02a} have shown that there exist nine families for pure states of four qubits (although there are infinite number of classes; one of these families depends on a continuous parameter). If one is given a single copy of a state and allowed to use SLOCC, Lo and Popescu \cite{ref:Lo&Popescu01} have shown that any pure two qubit state can be brought to a maximally entangled Bell state and Kent \emph{et al.} and others \cite{ref:Linden98,ref:Kent98,ref:Kent99,ref:Verstraete01a,ref:Verstraete02} have shown that the most entangled state one can obtain from a mixed state of two qubits is Bell diagonal. There has also been work on creating entanglement monotones by exploiting the $SL(2,\mathbb{C})$ invariance of certain entanglement measures \cite{ref:Teodorescu03,ref:Osterloh&Siewert05,ref:Osterloh08}.

In a lattice gauge theory the gauge field assigns a transformation to every pair of neighboring lattice points \cite{ref:Muenster&Walzl00}. This transformation is an element of the gauge group and is sometimes known as a parallel transporter; it parallel transports the property sitting at one lattice point to the neighboring point. For example, in lattice quantum chromodynamics, space-time is discretized into lattice points and the property sitting on the lattice points is color represented by a vector. To see how color changes from one lattice point to another one multiplies the vector by the parallel transporter. The assignment of parallel transporters to each link specifies the configuration of the gauge field. While the parallel transporters themselves change under gauge transformations, there is a natural gauge invariant observable on the lattice given by the trace of the total transformation around a loop (or plaquette). This is known as the Wilson loop. This total transformation is a measure of the curvature of the gauge field around that loop (equivalently the flux through the loop). The closer to the identity the transformation around the loop is, the less curved the gauge field is.

In this paper our $N$ lattice points are the $N$ qubits of a composite quantum state, the transformations between neighboring qubits are specified by the local operations that symmetrize the correlations in that two-qubit link and the gauge invariant observable, the Wilson loop, provides a measure of the degree of asymmetry of the bipartite correlations globally. We call this measure twist. These are correlations such as entanglement invariant under $SL(2,\mathbb{C})$, the gauge group of our theory.  In an earlier paper, Wootters explored a related idea and found evidence that a non-trivial twisting around a loop requires the sacrifice of some entanglement.
The definitions of twist that we introduce in the present paper are different from the one defined in the earlier paper
\cite{ref:Wootters02}, and they do not have the same interpretation.

The interpretation of our measure is as follows: Imagine taking a state comprised of $N$ qubits. One performs a local Kraus operation on the second qubit to symmetrize the correlations between the first and second qubits (symmetrize here means that the correlations specified by the expectation values of spin measurements on the first and second qubits labeled $1$ and $2$ are the same under interchange of the two qubits i.e. $\langle \sigma_i^1 \otimes \sigma_j^2 \rangle = \langle \sigma_j^1 \otimes \sigma_i^2 \rangle $). The indices $i$ and $j$ take the values $0,1,2,3$ with $\sigma_0=\mathbb{I}$ and $\sigma_{1}$, $\sigma_2$, $\sigma_3$ are the Pauli matrices. We can now imagine performing a local operation on the third qubit to symmetrize the correlations between qubits two and three, the link $2,3$. Now the links comprised of qubits $1,2$ and $2,3$ have been symmetrized. We can repeat this process along all links until the final one comprised of the last qubit, qubit $N$ and the first qubit. One now has a dilemma; the local operation performed on the first qubit that symmetrizes the link $N,1$ may cause the link $1,2$ to become asymmetric. The degree of mismatch between the initial and final local operations is the total transformation (also known as a holonomy) of an underlying correlation-induced gauge field around the loop and a measure of this mismatch is given by the trace of this total transformation, the Wilson loop. If one can simultaneously symmetrize all two qubit links in the loop, the overall transformation is trivial i.e. the identity. However, if this is not possible, the Wilson loop gives a degree of asymmetry in the bipartite correlations.

Even though our measure is defined for states of $N$ qubits, in this paper we restrict study to three qubit states, the most elementary loop one can conceive of. Explicitly, we show that all two qubit states have a trivial total transformation whereas all pure states of three qubits have a total transformation equal to a $\pi$ rotation. We then look at several examples of mixed states of three qubits. We provide examples of untwisted states and two examples of states with $SO(1,1)$ holonomy and link them with the concurrence in each link. These are examples we have been able to work out analytically since $SO(1,1)$ is a simple group depending on one parameter; it appears however, a generic mixed state of three qubits has the full $SO^+(1,3)$ structure depending on all six parameters. That is, a generic state cannot be untwisted. Although some analytical examples may be laborious to work out, one can calculate twist easily and quickly using numerical methods.

The ideas we outline here provide a novel bridge between the familiar gauge field theory approaches used in many areas of modern physics such as quantum field theory and the SLOCC protocols familiar in quantum information. The mathematical methods are the same although the meaning is very different. For example, the geometric phase \cite{ref:Berry84} has the same mathematical structure as our measure twist, however the holonomy is a result of the curvature of the space of quantum states or the parameter space of a Hamiltonian. In the case of twist the holonomy is induced by the bipartite correlations between quantum states. The key in the mapping of correlations to parallel transporters in this paper is the introduction of a Lorentz polar decomposition.

The paper is organized as follows: In the following section we show how one may assign a parallel transporter to a two qubit link and we introduce the measure twist. We then provide analytical results for states of two and three qubit states in sections~\ref{sec:results2} and \ref{sec:results3} respectively before concluding in section~\ref{sec:conclusions}.

\section{Mapping the correlations in a quantum state to a parallel transporter}

In this section we show how we map a parallel transporter to each two qubit link, the properties of these parallel transporters and the gauge invariance of a measure we call `twist', given by the Wilson loop.

The approach we use here is inspired by our previous work \cite{ref:Williamson10a} which showed how to generate local invariants. We represented bipartite states by correlation matrices, sometimes known as the Hilbert-Schmidt representation. That is, the two qubit density matrix, $\rho_{ab}$ is now represented by a real (although generally not positive or symmetric) four by four correlation matrix, $S(a,b)$ whose elements $i,j$ are given by
\begin{equation}
S(a,b)_{ij}=\tfrac{1}{2}\text{tr}[(\sigma_i^a\otimes\sigma_j^b) \rho_{ab}].
\end{equation}

We can also imagine performing local operations on the state $\rho_{ab}$ to take it to a new state $\rho_{ab}'= A\otimes B \rho_{ab} A^\dag \otimes B^\dag$. In the Hilbert-Schmidt representation, the local operations act on $S(a,b)$ as follows
\begin{equation}\label{eq:localoperations}
S(a,b)'=\mathcal{A}S(a,b)\mathcal{B}^T
\end{equation}
where the elements of the real, four by four matrices of the local operations on $a$ and $b$ in the new representation are given by
\begin{eqnarray}\label{eq:operators}
\mathcal{A}_{i_1 i_2}=\tfrac{1}{2}\text{tr}\left(A^\dag
{\sigma^a_{i_1}} A {\sigma^a_{i_2}}\right), \nonumber\\
{\mathcal{B}^T_{j_1 j_2}}=\tfrac{1}{2}\text{tr}\left(B
{\sigma^b_{j_1}} B^\dag {\sigma^b_{j_2}}\right).
\end{eqnarray}
The local operation $A$ must obey the constraint $A A^\dag \leq \mathbb{I}$ (and likewise for $B$) to take $\rho$ to a state with normalization $\leq 1$. However, the standard practice when thinking about SLOCC protocols is to rescale $A$ and $B$ by a positive constant so they have determinant one. That is, they are now elements of $SL(2,\mathbb{C})$ if they are reversible operations. In a SLOCC protocol one does not care about normalization, only that it may be possible to convert one state to another with some non-zero probability. This rescaling also makes sense when studying entanglement as the concurrence and three-tangle in a state remain invariant when considering local operations in this group as previously mentioned. From this point on we will restrict the local operations to be elements of $SL(2,\mathbb{C})$. When making this rescaling, the local operations in the Hilbert-Schmidt basis now become elements of $SO^+(1,3)$, the group of proper, orthochronous Lorentz transformations \cite{ref:Arrighi&Patricot03,ref:Verstraete01a}. This is due to the well known homomorphism $SL(2,\mathbb{C})\simeq SO^+(1,3)$. One can verify that $\mathcal{A}$ and $\mathcal{B}$ are indeed elements of $V\in SO^+(1,3)$ from the defining property of the Lorentz group; that it preserves the Minkowski metric, $\eta=\text{diag}\{1,-1,-1,-1\}$,
\begin{equation}\label{eq:Lorentzgroupproperty}
V \eta V^T=\eta.
\end{equation}
$\eta$ is equivalent to the spin-flip operator used in calculation of the well known entanglement measure, concurrence and is also equivalent to the singlet state $\ket{01}-\ket{10}$ when written as a correlation matrix.

\subsection{Lorentz polar decomposition and association of a parallel transporter to each two-qubit link}

Given a single copy of an arbitrary two-qubit state what is the most entangled state one can convert it to using SLOCC? In a series of papers \cite{ref:Linden98,ref:Kent98,ref:Kent99,ref:Verstraete01a,ref:Verstraete02} it was shown that the most entangled state one can produce is a Bell diagonal state, a mixture of some or all of the four Bell states $\ket{\Psi^\pm}=\ket{01}\pm\ket{10}$, $\ket{\Phi^\pm}=\ket{00}\pm\ket{11}$. In the correlation matrix representation, the four Bell states are the diagonal matrices
\begin{eqnarray}
S_{\ket{\Psi^-}}&=&\tfrac{1}{2}\text{ diag}\{1,-1,-1,-1\},\nonumber\\
S_{\ket{\Psi^+}}&=&\tfrac{1}{2}\text{ diag}\{1,1,1,-1\},\nonumber\\
S_{\ket{\Phi^-}}&=&\tfrac{1}{2}\text{ diag}\{1,-1,1,1\},\nonumber\\
S_{\ket{\Phi^+}}&=&\tfrac{1}{2}\text{ diag}\{1,1,-1,1\}.
\end{eqnarray}
Since a Bell diagonal state is just a mixture of these pure Bell state correlation matrices, a Bell diagonal state must also be a diagonal correlation matrix. This prompts a natural decomposition of an arbitrary two-qubit state represented by a correlation matrix $S(a,b)$ \cite{ref:Verstraete01a}:
\begin{equation}\label{eq:LSVD}
S(a,b)=V_a \Sigma_{ab} W_b^T,
\end{equation}
where $\Sigma_{ab}=\text{diag}\{s_0,s_1,s_2,s_3\}$ represents a Bell diagonal state and $V_a$ and $W_b$ (elements of $SO^+(1,3)$) are the local operations taking one from the Bell diagonal state to the state represented by $S(a,b)$. This is a special form of a singular value decomposition (SVD) with a SLOCC operational interpretation. With the usual SVD one makes the entries of $\Sigma$, the singular values, real and positive and puts them in non-increasing order down the diagonal. Since $V,W \in SO^+(1,3)$ we can order the Lorentz singular values in non-increasing order and they are real since $S$, $V$ and $W$ are real. However, we cannot generally make them all positive since doing so could result in a density matrix with negative probabilities, an invalid physical state. The leading Lorentz singular value, $s_0$, is always positive and the largest (it represents the normalization of the state). The remaining three Lorentz singular values $s_1$, $s_2$ and $s_3$ take the same sign (generically negative). The ordering of these remaining three singular values will not affect the parallel transporter we assign to each link, however the sign they take will. These singular values represent the expectation values of spin in the three spatial directions $s_i=s_0\langle \sigma_i \otimes \sigma_i\rangle$ rescaled by the normalization of the state $s_0$.

The Lorentz singular values are $SL(2,\mathbb{C})$ invariants and are closely related to the concurrence in the two-qubit state \cite{ref:Kent99,ref:Verstraete01a}. That is, the entanglement in the state is related to the (dilated) magnitude of the correlations. It is a Lorentz scalar.
The concurrence $\mathcal{C}$ for an arbitrary two qubit state is
\begin{equation}
\mathcal{C}=\max\{0,-\text{tr}\Sigma\}.
\end{equation}

By rewriting the Lorentz SVD, we can also find a left Lorentz polar decomposition. That is, we may re-write $S$ as
\begin{equation}
S(a,b)=\Lambda(a,b) \tilde{S}(a,b)
\end{equation}
where $\Lambda(a,b)=V_a \eta W_b^T \eta \in SO^+(1,3)$ and $\tilde{S}(a,b)= W_b \Sigma_{ab} W_b^T$. This is equivalent to our previous decomposition of $S$ except now $\tilde{S}(a,b)$ represents a state with correlations that are symmetric under interchange of the qubits and $\Lambda(a,b)$ represents the local operation performed on $a$ that symmetrizes the correlations. Moreover if $S$ is full rank, $\Lambda$ is unique although $V$ and $W$ may not be. $\Lambda$ is the parallel transporter we assign to each two-qubit link in our gauge theory. We give more details about the properties of $\tilde{S}$ and $\Lambda$ and their calculation in sections~\ref{sec:properties} and \ref{sec:calculation} respectively.

Just as one can also decompose an arbitrary matrix into the right polar form using the standard SVD, one can also do the same for the Lorentz SVD. That is, we may write $S(a,b)=\tilde{S}(a,b)'\Lambda(a,b)'$ where $\tilde{S}(a,b)'=V_a \Sigma_{ab} V_a^T$ and $\Lambda(a,b)'=\eta V_a \eta W_b^T$. The interpretation of this decomposition is that now $\Lambda(a,b)'$ represents the local operation performed on qubit $b$ that symmetrizes the correlations in the two qubit state. $\tilde{S}(a,b)'$ is the (different) symmetrized state. While $\Lambda(a,b)'\neq \Lambda(a,b)$ (they are related by $\Lambda(a,b)'=\eta\Lambda(a,b)\eta$) the eigenvalues and trace of a cyclic product of the $\Lambda$ representing the overall transformation around a loop remain the same provided one is consistent in taking just the left or the right polar decomposition. It is this total transformation that we look at in the next section.

One might worry that restricting the matrices $V$ and $W$ to be elements of the Lorentz group, one may not always be able to bring a two qubit correlation matrix to Bell diagonal form and indeed some cases do exist. These non-diagonalizable cases are not generic however. Verstraete \emph{et al.} \cite{ref:Verstraete02} have listed the non-diagonalizable cases, most of which turn out to be product states comprised of one or more pure subsystems. We do not consider two qubit links composed of product states since it seems unlikely one can consistently define a correlation related parallel transporter in such examples. There is one other non-diagonalizable case, it is the state $p_0\matx{00}+p_1\matx{\Psi^+}+p_2\matx{\Psi^-}$. These states can still be brought to a well defined diagonal form with infinite Lorentz transformations, a process known as quasi-distillation, and we use this technique in section~\ref{sec:results3}. However, one does not have to apply infinite Lorentz transformations to assign a parallel transporter to these states; the parallel transporter only has to symmetrize ($S=S^T$) the correlations in the two qubit link. This can be done without the use of infinite Lorentz transformations.

\subsection{Twist as a measure of asymmetry of correlations: A Lorentz group gauge theory}

We now take a cyclic product of the parallel transporters $\Lambda$ representing a loop around the lattice points (qubits) to form our measure of the degree of asymmetry of bipartite correlations globally. We define this measure, twist $\xi$ as
\begin{equation}
\xi(ab\cdots z)=\tfrac{1}{4}\text{tr}\{\Lambda(a,z)\cdots\Lambda(c,b)\Lambda(b,a)\}.
\end{equation}
The constant $1/4$ is chosen to give $\xi=1$ when the overall transformation is the identity. We now ask how the parallel transporters transform under local operations or in the language of gauge theories, gauge transformations $U \in SO^+(1,3)$. From Eqs.~(\ref{eq:localoperations}) and (\ref{eq:LSVD}) we see that $V_a \rightarrow U_a V_a$ and $W_b^T \rightarrow W_b^T U_b^T$. This implies $\Lambda(a,b)$ transforms as
\begin{equation}
\Lambda(a,b)\rightarrow U_a \Lambda(a,b) U_b^{-1},
\end{equation}
just as parallel transporters should indeed transform under gauge transformations. We have used the identities $\eta \eta=\mathbb{I}$ and $\eta U^T \eta = U^{-1}$. One can now see under gauge transformations, the parallel transporters do change but the total transformation around a loop only changes up to a similarity transformation leaving the trace and the eigenvalues of the overall transformation $\Lambda(a,z)\cdots\Lambda(c,b)\Lambda(b,a)$ invariant under $SO^+(1,3)$. This overall transformation is sometimes also known as a holonomy. In the case of twist it is a Lorentz group holonomy.

\begin{figure}
\begin{center}
\subfigure[]{\includegraphics[width=3.5cm]{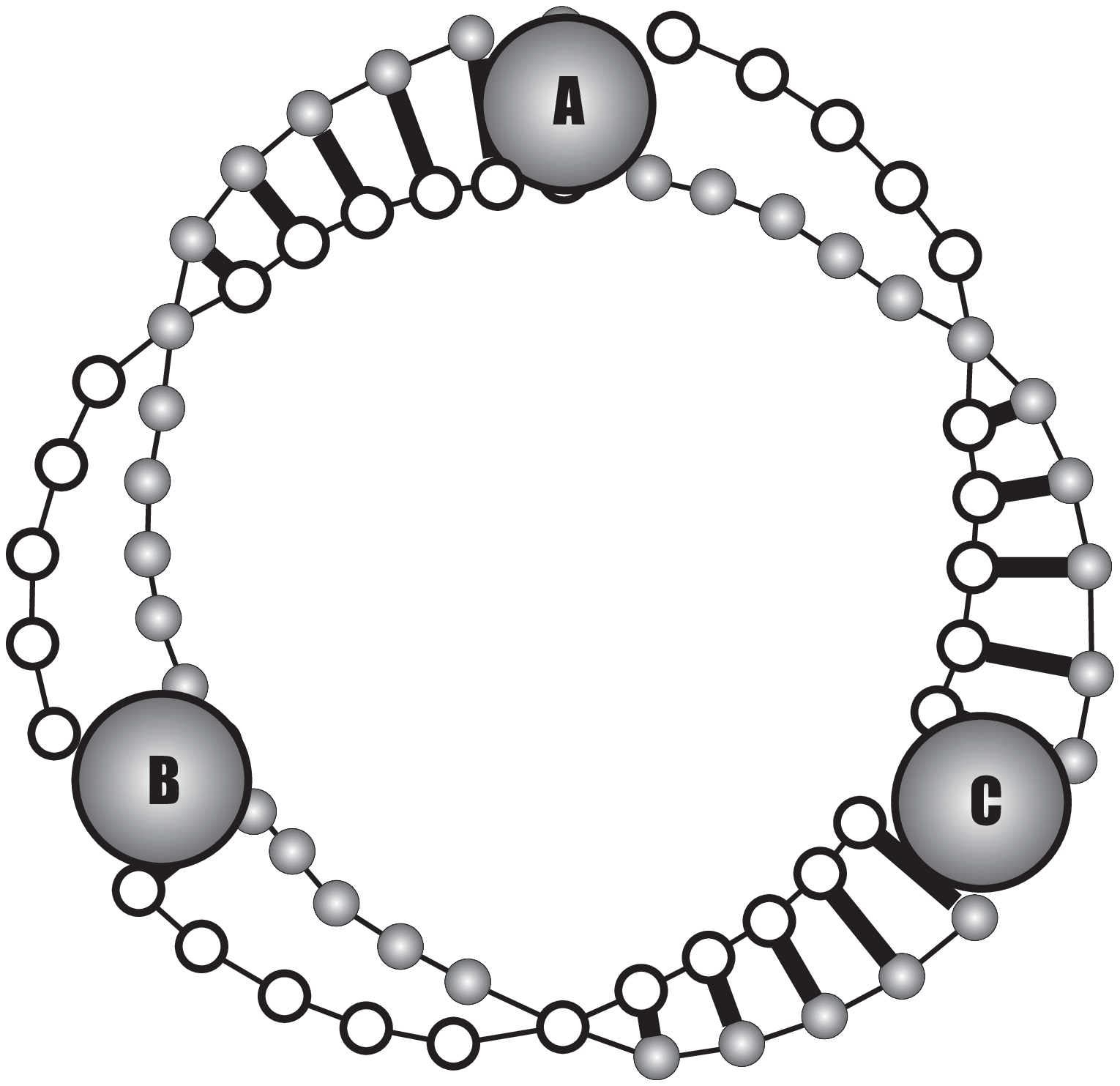}}
\hspace{1cm}
\subfigure[]{\includegraphics[width=3.5cm]{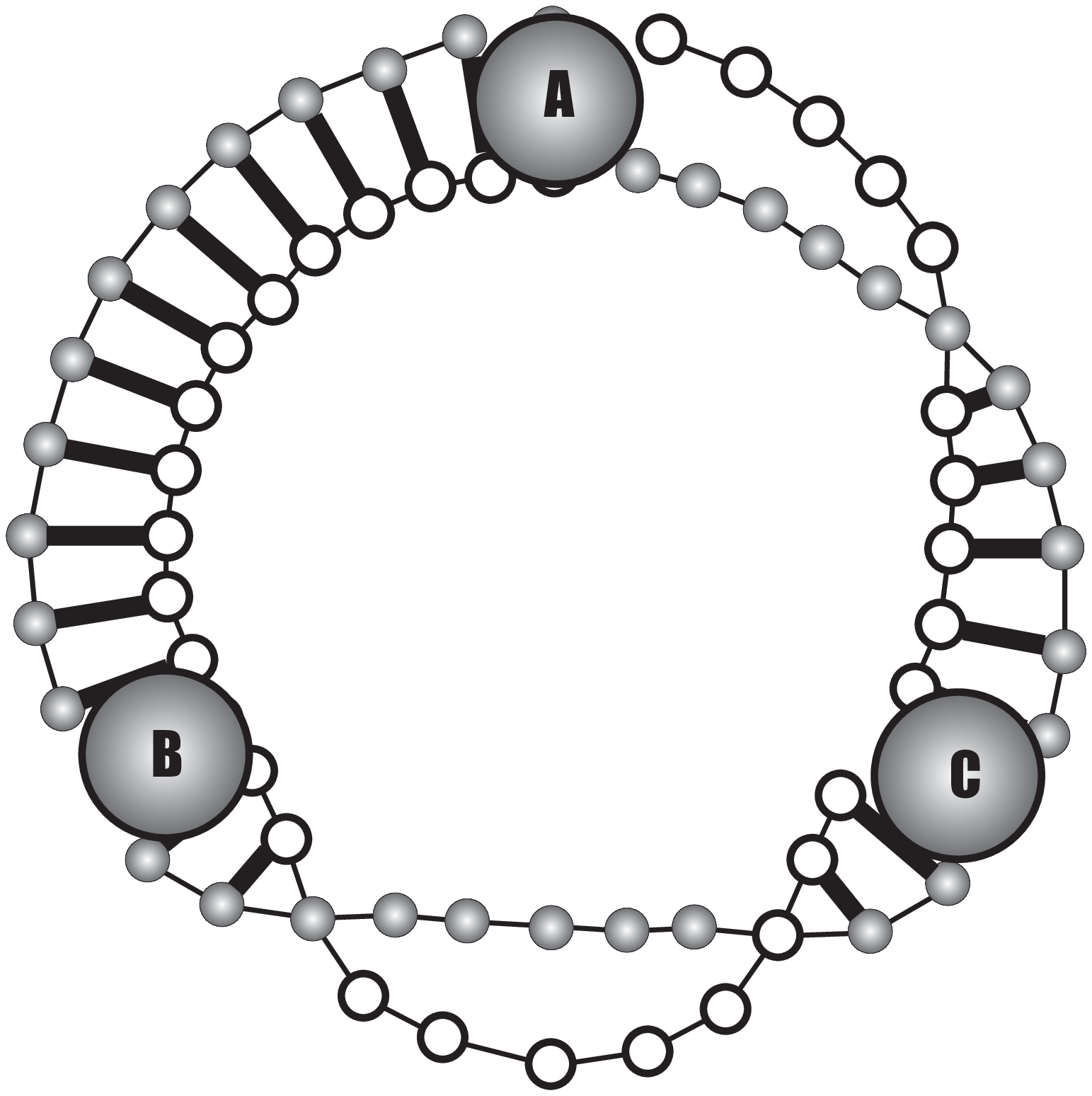}\label{sec:SL2invariants:fig:twist2}}
\hspace{1cm}
\subfigure[]{\includegraphics[width=3.5cm]{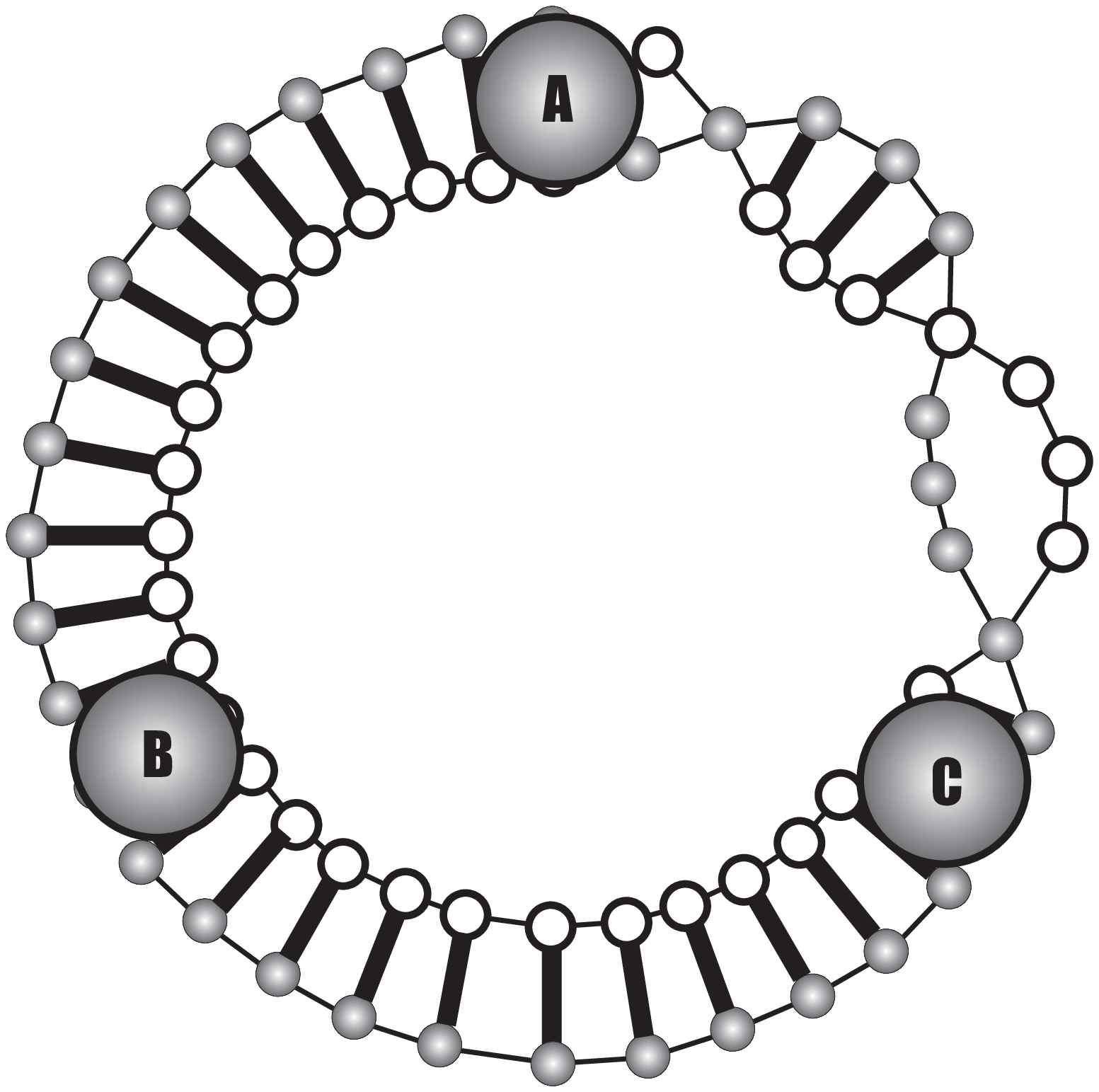}\label{sec:SL2invariants:fig:twist3}}
\end{center}
\caption[Graphical representation of twist for
three qubits.]{Graphical representation of twist $\xi(abc)$ for a state of three qubits. The ribbons between
the qubits represent the `twist', the degree of asymmetry in the correlations in each two-qubit link. The
twist in each link is not invariant by itself however. One can
untwist each link by careful choice of local gauge transformation (representing a local operation)
$U \in SO^+(1,3)$. Fig.~\ref{sec:SL2invariants:fig:twist2}
corresponds to the local transformation on $b$ that results in the
link $\Lambda(b,a)$ being the identity $\Lambda(b,a)\rightarrow
U_b \Lambda(b,a)$. Similarly fig.~\ref{sec:SL2invariants:fig:twist3}
corresponds to the local transformation on qubit $c$ that untwists
$\Lambda(c,b)U_b^{-1}$ i.e.
$\Lambda(c,b)U_b^{-1}\rightarrow U_c \Lambda(c,b)U_b^{-1}$.
One can still perform a local transformation on qubit $a$ however
the total twist in the link will not change. Even though locally we
can untwist each link we cannot gauge away the twist
globally.}\label{fig:twist}
\end{figure}

Twist is therefore a measure of the asymmetry of the correlations you cannot `gauge' away globally. One is reminded of the Aharonov-Bohm effect \cite{ref:Aharonov&Bohm59} and the geometric phase \cite{ref:Berry84} as illustrations of features that although can be gauged away locally, cannot be gauged away globally. Twist shares another property of the Aharonov-Bohm effect; that one cannot associate a twist to an individual link since each parallel transporter associated to that link can be `gauged away'. We can only meaningfully associate a twist globally to the entire loop of links. In this sense, twist is a nonlocal property of the bipartite correlations present in the loop. We give a simple illustration of the idea in Fig.~\ref{fig:twist}.

% put brief details about experimental vertification
Experimentally determining the amount of twist in an unknown quantum state is a similar problem to determing the amount of entanglement. Both can be interferred from state tomography. If however, one \emph{knows} the state beforehand, one can confirm the amount of twist by applying sequentially local operations to each qubit around a particular loop that symmetrize or `untwist' the bipartite links. Using our three qubit example illustrated in figure~\ref{fig:twist}, one applies the local operation $\Lambda(b,a)^{-1}$ to qubit $b$ to symmetrize the link $ba$ and then applies $\Lambda(b,a)^{-1}\Lambda(c,b)^{-1}$ to qubit $c$ to symmetrize $bc$. One should find that the last link $ac$ can only be symmetrized by the overall transformation $\Lambda(b,a)^{-1}\Lambda(c,b)^{-1}\Lambda(a,c)^{-1}$ on qubit $a$ but in doing so makes the neighboring link, $ab$, assymetric by the same amount. This could be confirmed by tomography on one of the bipartite links. One should note since these operations are generally probabilistic they are not always successful. If one obtains the undesired outcome one would need to apply a compensating operation that cancels out this undesired outcome which is also generally probabilistic.

%Experimental verification details... Of course one can always verify this stuff via state tomography and I don't see a way around doing state tomography at some point to verify you have a symmetric state. Are there more efficient ways of doing the verification? Think about how one measures entanglement in an experimental situation. Should also mention this is all to do with bipartite correlations i.e. if you include the third qubit and do local operations on that qubit it can affect the state nonlocally through the three- or more-party correlations. Note that the AB effect and the geometric phase are dynamic or 'kinematic' effects i.e. you put the state through some sort of evolution. In what we have here, everything is static. No dynamics. Need to watch what we call multiparty SLOCC invariant here too.

\subsection{Properties of $\Lambda$ and $\tilde{S}$}\label{sec:properties}

In the Lorentz SVD given in eq.~(\ref{eq:LSVD}) we have chosen to make the three singular values $s_1$, $s_2$ and $s_3$ take the same sign as this results in a unique $\Lambda$ when the singular values are all non-zero. Giving them the same sign also seems natural as the three spatial correlation singular values are treated equally. There are two possible cases: (i) The sign is negative if $\det S<0$ and (ii) the sign is positive if $\det S>0$. For case (i), the diagonal matrix $\Sigma$ represents a Bell state mixture, the largest component of this mixture being the singlet, $\ket{\Psi^-}$. In case (ii), the singlet makes up the smallest component in the Bell diagonal mixture. In both cases the ordering of the remaining three Bell states, $\ket{\Phi^-}$, $\ket{\Phi^+}$ and $\ket{\Psi^+}$ does not affect the form of the parallel transporters $\Lambda$ however it does affect the form of the symmetrized correlation matrix $\tilde{S}$. One can see why the ordering of the singlet is the determining factor on the form of $\Lambda$ from its special properties, namely that it is invariant under symmetric local operations i.e. $A \otimes A \ket{\Psi^-}=\ket{\Psi^-}$ where $A \in SL(2,\mathbb{C})$. This last fact can be seen even more clearly in the correlation matrix form since $S_{\ket{\Psi^-}}\equiv \eta$ and our relation becomes equivalent to the defining property of the Lorentz group, eq.~(\ref{eq:Lorentzgroupproperty}). That is, under symmetric local operations, the singlet state is invariant, it remains in the anti-symmetric state space, a state space consisting of a single point. To change the ordering of the singlet state in the Bell mixture one has to apply different local operations to the two qubits. The three remaining triplet Bell states are not invariant under symmetric, local operations, however they do of course remain in the much larger symmetric state space. In fact one can transform any triplet Bell state into any other using a subset of these symmetric local operations. Using these facts we can see that the ordering of the triplet states in a Bell state mixture does not affect the form of $\Lambda$ since making the replacements $V\rightarrow V U$ and $W\rightarrow W U$ leave it invariant. $U \in SO^+(1,3)$ is a local operation applied to both subsystems. One also notes that the absolute singlet content in $\tilde{S}$ is unaffected by symmetric local operations. It is thus the ordering of the singlet that determines the form of the parallel transporters and the absolute singlet content in the symmetrized states.

The parallel transporters are well defined and unique when $S$ is full rank however they are not unique when one or more $s_i$ are zero. It is necessary but not sufficient for $S$ to be full rank for entanglement to exist in that link. Therefore the assignment of a unique parallel transporter to a link does not mean that link is entangled, however it does imply that the correlations in that link are `quantum'. That is, correlations that do not appear in classical physics. This type of weaker correlation is generally known in the literature under the name discord \cite{ref:Olivier&Zurek02}. In this paper we only consider the cases where $S$ is full rank and therefore the parallel transporters are unique. However one may be able to define the notion of a partial holonomy in the cases of $\text{rank}(S)\geq 2$ corresponding to classically correlated (rank 2) and partially quantum correlated (rank 3), by assigning a parallel transporter over the subspace in which the singular values are non-zero. This idea has been proposed in the context of geometric phases by Kult \emph{et al.} \cite{ref:Kult06}. In the case $S$ is rank 1, the bipartite link is a product state meaning that the two qubits are uncorrelated. The two qubits know nothing about each other and never can do if we restrict to using only local operations. Assigning a parallel transporter in this case does not seem possible.

%Invariance of $\Lambda$ under symmetric operations, invariance of $\ket{\Psi^-}$ under symmetric operations. What the signature means in terms of Bell state decomposition.
%
%Invariance of singlet under symmetric transformations and the transformation of one into the other of triplets under symmetric transformations. Symmetric transformations do not alter the parallel transporters.
%
%Symmetrization of correlations - there are four possible states corresponding to different amounts of singlet. Choose the one with the maximal amount of singlet.
%
%Extra parallel transport rules in the case $S$ is not full rank? What does full rank imply? Talk about existence of quantum correlations, not necessarily entanglement.
%
%Those cases where you have product states or $S$ is not full rank. What to write here.

\subsection{Calculation of the parallel transporters, $\Lambda$}\label{sec:calculation}

Calculation of a parallel transporter for a given link involves finding the two elements, $V$ and $W$ in $SO^+(1,3)$ that diagonalize $S$ in eq.~(\ref{eq:LSVD}). A typical $S$ will have all non-zero elements. An arbitrary element of $SO^+(1,3)$ may be decomposed into a product of two different elements of $R_1,R_2 \in SO(3)$, two spatial rotations, sandwiching an element of $B\in SO(1,1)$, a Lorentz boost along one spatial axis, $V=R_1 B R_2$. Explicitly the $4 \times 4$ representations are:
\begin{eqnarray}
R_i&=&\begin{pmatrix}
  1 & 0 & 0 & 0 \\
  0 & c_{\alpha} c_\beta & s_\alpha c_\beta s_\gamma - s_\beta c_\gamma & s_\alpha c_\beta c_\gamma + s_\beta s_\gamma \\
  0 & c_\alpha s_\beta & s_\alpha s_\beta s_\gamma + c_\beta c_\gamma & s_\alpha s_\beta c_\gamma - c_\beta s_\gamma \\
  0 & -s_\alpha & c_\alpha s_\gamma & c_\alpha c_\gamma \\
\end{pmatrix},\\
B&=&\begin{pmatrix}
    \cosh \varphi & 0 & 0 & \sinh \varphi \\
    0 & 1 & 0 & 0 \\
    0 & 0 & 1 & 0 \\
    \sinh \varphi & 0 & 0 & \cosh \varphi \\
  \end{pmatrix},
\end{eqnarray}
where $c_\alpha=\cos\alpha$ and $s_\alpha=\sin\alpha$ and analogous relations for the two other Euler angles $\beta$ and $\gamma$. This decomposition suggests an iterative procedure to diagonalize $S$ by applying a rotation $R$ followed by a boost $B$ to each qubit in the link. The aim is to depolarize each of the individual qubit's Bloch vectors simultaneously making each individual qubit's density matrix proportional to the identity. Rotations do not change the length of these Bloch vectors but they can change each of the spatial components. One strategy one might try is to first rotate each qubit's Bloch vector so that it is aligned along one spatial axis, say the $z$ axis, making each of the entries $S_{01}$, $S_{02}$ and $S_{10}$, $S_{20}$ zero and then boost along the $z$ axis in the opposite direction to which that Bloch vector points by a small amount. One repeats this until both qubits are simultaneously depolarized. The state is then proportional to a Bell diagonal state, although $S$ may not be diagonal or in the correct signature. One can diagonalize the final $3\times 3$ block by applying two final spatial rotations, $R_1$. More detailed calculational techniques are reported in \cite{ref:Kent99,ref:Verstraete01a,ref:Verstraete03,ref:Cen02}.

Another strategy rather than finding $V$ and $W$ directly is to simplify the form of the two qubit links by applying gauge transformations to each of the qubits since the overall transformation around a loop is gauge invariant. For example in the three qubit case, one may simplify calculation by choosing $U_a$, $U_b$ and $U_c$ to reduce the complexity of diagonalizing $S$ i.e. make the transformations
\begin{eqnarray}
S(b,a)&\rightarrow& U_b S(b,a) U_a^T, \nonumber\\
S(c,b)&\rightarrow& U_c S(c,b) U_b^T, \nonumber\\
S(a,c)&\rightarrow& U_a S(a,c) U_c^T.
\end{eqnarray}

One may also find it simpler to work in the density matrix basis; rather than dealing with real $4\times 4$ matrices, one can find $2 \times 2$ complex matrices, $A,B \in SL(2,\mathbb{C})$, that take each two qubit density matrix, $\rho$, to a Bell diagonal state, $\rho'$. That is $\rho'=A\otimes B \rho A^\dag \otimes B^\dag$.
%How to calculate the $\Lambda$, either numerically or analytically. Use of gauge transformations or straight calculation. $SL(2,\mathbb{C})$ form or $SO^+(1,3)$ form. Decomposition of $SO^+(1,3)$ into $SO(1,1)$ and two copies of $SO(3)$. References for calculation and tips \cite{ref:Cen02}.

Analytical calculation of $V$ and $W$ is possible when the singular values are non-degenerate corresponding to unique $V$ and $W$. One can solve the eigenproblem for $S^T \eta S \eta = W\Sigma^2 W^{-1}$. $W^{-1}$ is a Lorentz matrix whose columns are the eigenvectors of $W$. Likewise one can find $V$ by solving the eigenproblem for $S\eta S^T\eta$.

\section{Results for two-qubit states}\label{sec:results2}

\subsection{Pure states of two qubits}
We first consider a pure state of two qubits which can be written (up to local unitary equivalence) as $\alpha\ket{00}+\beta\ket{11}$ where $\alpha$ and $\beta$ are real. We wish to calculate twist for this state, the path from qubit $a$ to $b$ and back again given by
\begin{equation}
\xi(ab)=\tfrac{1}{4}\text{tr}\{\Lambda(a,b)\Lambda(b,a)\}.
\end{equation}
We can use gauge transformations $U_a$ and $U_b$ to simplify the calculation of $\Lambda$. Using the $SO^+(1,3)$ transformation on qubit $a$
\begin{equation}
U_a=\begin{pmatrix}
      \tfrac{1}{2\alpha \beta} & 0 & 0 & \tfrac{\beta^2-\alpha^2}{2\alpha\beta} \\
      0 & -1 & 0 & 0 \\
      0 & 0 & 1 & 0 \\
      \tfrac{\alpha^2-\beta^2}{2\alpha\beta} & 0 & 0 & -\tfrac{1}{2\alpha\beta} \\
    \end{pmatrix}
\end{equation}
and the transformation $U_b=\mathbb{I}$ one can convert each link to $S(a,b)=\Sigma_{ab}=\alpha\beta \times \text{diag}\{1,-1,-1,-1\}$, a state proportional to the singlet, which, of course, has symmetric correlations under interchange of the qubits. This is the well known result that given a single copy of a pure, entangled state of two qubits, one can always convert it to a singlet using SLOCC \cite{ref:Lo&Popescu01}. The gauge transformation $U_a$ is equivalent to the $SL(2,\mathbb{C})$ matrix
\begin{equation}
A=\begin{pmatrix}
  0 & -\sqrt{\tfrac{\alpha}{\beta}} \\
  \sqrt{\tfrac{\beta}{\alpha}} & 0 \\
\end{pmatrix}
\end{equation}
in the density matrix basis.

Since we can gauge away all the freedom in each link, the parallel transporters can all be consistently gauge transformed to $\Lambda(a,b)=\Lambda(b,a)^{-1}=\mathbb{I}$ giving $\xi(ab)=1$. This gives the result that all pure states of two qubits are untwisted.

\subsection{Mixed states of two qubits}

Arbitrary two qubit states are also untwisted ($\xi(ab)=1$). This is simply seen by noting that $S(a,b)=S(b,a)^T$. Since $S(a,b)=V_a \Sigma_{ab} W_b^T$, $S(b,a)=W_b \Sigma_{ab} V_a^T$. Note $\Sigma_{ab}$ is now generally no longer proportional to the singlet but it does have the same signature $(+,-,-,-)$. The decomposition of $S(a,b)$ implies $\Lambda(a,b)=\Lambda(b,a)^{-1}$ or equivalently there exist gauge transformations, $U_a$ and $U_b$ that untwist the link. For example, choose $U_a=V_a^{-1}$ and $U_b=W_b^{-1}$. This illustrates another key property of gauge theories; one can only get the possibility of a non-trivial holonomy from a loop that encloses some area. Clearly for two-qubit paths this can never be done. One needs to consider states of three qubits and higher for the possibility to see non-trivial twist.

\section{Results for three qubit states}\label{sec:results3}

\subsection{Pure states of three qubits}

It is well known that a pure state of three qubits can be brought to one of two standard forms using SLOCC: (i) The GHZ state $\ket{000}+\ket{111}$ or (ii) the W state $\ket{001}+\ket{010}+\ket{100}$. A generic pure three qubit state is of the GHZ class \cite{ref:Dur00}. Since both of these states are symmetric under permutations of the qubits we expect that pure states of three qubits have trivial twist. However, we prove that the overall transformation is a $\pi$ rotation.

\emph{\textbf{Theorem}: A generic state of three qubits in a pure state has $\xi(abc)=0$, with total transformation $\Lambda(a,c)\Lambda(c,b)\Lambda(b,a)\equiv\text{diag}\{1,1,-1,-1\}$ in some basis. That is, the twist around the three qubit loop is a $\pi$ (Pauli) rotation around some spatial axis.}

\emph{\textbf{Proof:}} We prove this by again making a careful choice of gauge transformation, $U$, on each qubit locally. After this choice one finds that there is a Pauli rotation that cannot be gauged away globally.

We use parameterizations of GHZ and W states first given by D\"{u}r et al. \cite{ref:Dur00} and solve for each class individually. These two parameterizations populate the whole of pure three qubit state space.

Any state in the SLOCC GHZ class can be written as
\begin{equation}
\ket{GHZ}=c_\delta\ket{000}+s_\delta e^{i \varphi}\ket{\phi_a \phi_b \phi_c}
\end{equation}
up to local unitaries. In the GHZ form $\ket{\phi_a}=c_\alpha\ket{0}+s_\alpha\ket{1}$, $\ket{\phi_b}=c_\beta\ket{0}+s_\beta\ket{1}$ and $\ket{\phi_c}=c_\gamma\ket{0}+s_\gamma\ket{1}$. $c$ and $s$ stand for $\cos$ and $\sin$ respectively and the real angles have the ranges $\alpha,\beta,\gamma \in (0,\pi/2]$, $\delta \in (0,\pi/4]$ and $\varphi \in (0,2\pi]$.

We next calculate $S(b,a)$, $S(c,b)$ and $S(a,c)$ for each two qubit link and try to get each link into the $\Sigma$ form using local operations (gauge transformations) $U$. We choose the following $SL(2,\mathbb{C})$ transformations
\begin{eqnarray}
%A&=&\begin{pmatrix}
%     \left(\frac{e^{i \phi} c_{\alpha}^{2}t_\delta}{t_\alpha^2}\right)^{\frac{1}{4}} & -\left(\frac{e^{-i \phi}}{s_\alpha^2 t_\delta}\right)^{\frac{1}{4}} \\
%     \left({e^{i \phi} s_{\alpha}^2 t_\delta}\right)^{\frac{1}{4}} & 0 \\
%  \end{pmatrix}
%\\
%B&=&\begin{pmatrix}
%     \left(\frac{e^{-i\phi}}{s_\beta^2 t_\delta}\right)^{\frac{1}{4}}& \left(\frac{e^{i\phi} c_\beta^2 t_\delta}{t_\beta^2}\right)^{\frac{1}{4}}\\
%    0 & \left( e^{i\phi} s_\beta^2 t_\delta\right)^{\frac{1}{4}} \\
%  \end{pmatrix}
%\\
%C&=&\begin{pmatrix}
%    \left(\frac{e^{i\phi} c_\gamma^2 t_\delta}{t_\gamma^2}\right)^{\frac{1}{4}}& -\left(\frac{e^{-i\phi}}{s_\gamma^2 t_\delta}\right)^{\frac{1}{4}}\\
%   \left(e^{i\phi} s_\gamma^2 t_\delta\right)^{\frac{1}{4}} & 0 \\
% \end{pmatrix}
%
A&=&\begin{pmatrix}
      0 & \left(\frac{e^{-i \varphi}}{s_\alpha^2 t_\delta}\right)^{\frac{1}{4}}\\
     -\left({e^{i \varphi} s_{\alpha}^2 t_\delta}\right)^{\frac{1}{4}} & \left(\frac{e^{i \varphi} c_{\alpha}^{2}t_\delta}{t_\alpha^2}\right)^{\frac{1}{4}} \\
  \end{pmatrix},
\\
B&=&\begin{pmatrix}
     \left( e^{i\varphi} s_\beta^2 t_\delta\right)^{\frac{1}{4}} & -\left(\frac{e^{i\varphi} c_\beta^2 t_\delta}{t_\beta^2}\right)^{\frac{1}{4}}\\
    0 & \left(\frac{e^{-i\varphi}}{s_\beta^2 t_\delta}\right)^{\frac{1}{4}} \\
  \end{pmatrix},
\\
C&=&\begin{pmatrix}
    0 & \left(\frac{e^{-i\varphi}}{s_\gamma^2 t_\delta}\right)^{\frac{1}{4}}\\
   -\left(e^{i\varphi} s_\gamma^2 t_\delta\right)^{\frac{1}{4}} & \left(\frac{e^{i\varphi} c_\gamma^2 t_\delta}{t_\gamma^2}\right)^{\frac{1}{4}} \\
 \end{pmatrix}.
\end{eqnarray}
$t$ here stands for $\tan$ and the corresponding Lorentz element $U_a$ is given either by Eq.~(\ref{eq:operators}) or by
\begin{equation}\label{eq:Voperator}
U_a=T(A\otimes A^*)T^\dag
\end{equation}
and likewise for $U_b$ and $U_c$. $T$ is the matrix
\begin{equation}
T=\frac{1}{\sqrt{2}}\begin{pmatrix}
    1 & 0 & 0 & 1 \\
    0 & 1 & 1 & 0 \\
    0 & i & -i & 0 \\
    1 & 0 & 0 & -1 \\
  \end{pmatrix}.
\end{equation}
Using the gauge transformations $U_a$, $U_b$ and $U_c$ on $S(b,a)$, $S(c,b)$ and $S(a,c)$ i.e.
\begin{eqnarray}
S(b,a) \rightarrow U_b S(b,a) U_a^T, \nonumber\\
S(c,b) \rightarrow U_c S(c,b) U_b^T, \nonumber\\
S(a,c) \rightarrow U_a S(a,c) U_c^T,
\end{eqnarray}
we can reduce each $S$ to diagonal form. However, the gauge transformed $U_a S(a,c)U_c^T$ has the signature $(+,+,-,+)$.  We therefore need to multiply it by $\text{diag}\{1,-1,1,-1\}$, a Pauli $y$ rotation, to put the final link in the desired form. This final Pauli rotation is the holonomy, since we cannot gauge it away. In other words
\begin{equation}
\Lambda(a,c)\Lambda(c,b)\Lambda(b,a)=\text{diag}\{1,-1,1,-1\}
\end{equation}
giving $\xi(abc)=0$ for all states in the GHZ SLOCC class.

Note that the result would have been the same without the use of the gauge transformations since the overall transformation is gauge invariant. The careful choice of gauge transformation just made the calculation simpler.

Any state in the SLOCC W class can be written as
\begin{equation}\label{eq:wstate}
\ket{W}=w\ket{000}+x\ket{001}+y\ket{010}+z\ket{100}
\end{equation}
up to local unitaries with $w,x,y,z\geq 0$. We can follow the same strategy as for the GHZ case, trying to reduce the complexity of finding $\Lambda$ by gauging away as much of the twist as possible. If we choose the gauge transformations
\begin{eqnarray}
A&=&Q\begin{pmatrix}
      \sqrt{\frac{z}{x}} & -\frac{w}{\sqrt{xz}} \\
      0 & \sqrt{\frac{x}{z}} \\
    \end{pmatrix},
\\
B&=&Q\begin{pmatrix}
      -i\sqrt{\frac{y}{x}} & 0 \\
      0 & i\sqrt{\frac{x}{y}} \\
    \end{pmatrix},
\\
C&=&Q,
\end{eqnarray}
we can reduce $S(b,a)$ and $S(c,b)$ to the desired $\Sigma$ form. $S(a,c)$ is also diagonal but with the wrong signature $(+,+,+,-)$. Again, we have given these transformations in $SL(2,\mathbb{C})$ form. However, it is simple to find $U_a$, $U_b$ and $U_c$ in $SO^+(1,3)$ form using either Eq.~(\ref{eq:operators}) or Eq.~(\ref{eq:Voperator}). A key step in the gauge transformation procedure is using the local operator $Q=\text{diag}\{n,1/n\}$. To get these states in the desired Bell diagonal form we must take the limit $n\rightarrow \infty$ which takes $Q$ to the limit of a projective measurement. Since projections are not invertible they are not group elements and strictly do not belong to representations of $SL(2,\mathbb{C})$ or $SO^+(1,3)$. However, one can get infinitesimally close to a projective measurement and the Lorentz singular values of $S$ are well defined in this limit. In fact each $S$ becomes proportional to a pure Bell state, the links $S(b,a)\rightarrow yz\times\text{diag}\{1,-1,-1,-1\}$ and $S(c,b)\rightarrow xy\times\text{diag}\{1,-1,-1,-1\}$ being proportional to the singlet, $\ket{\Psi^-}$ and the final link $S(a,c)\rightarrow xz\times\text{diag}\{1,1,1,-1\}$ being proportional to $\ket{\Psi^+}$. This is a process known as `quasi-distillation' \cite{ref:Horodecki99}. Two qubit states of this form can be brought to pure Bell states but the probability of success decreases to zero.

These gauge transformations untwist all the links except for $S(a,c)$ which has the incorrect signature. Therefore the overall holonomy for W states is again a Pauli rotation, $\Lambda(a,c)\Lambda(c,b)\Lambda(b,a)=\text{diag}\{1,-1,-1,1\}$ ($z$ rotation in this particular basis) giving $\xi(abc)=0$ once again.

\emph{\textbf{Discussion:}} We find that the total transformation around a loop for pure states of three qubits is equivalent to the matrix $\text{diag}\{1,-1,1,-1\}$. In the GHZ case we find that the links $ab$ and $bc$ could be brought to the form $\lambda_0\matx{\Psi^-}+\lambda_1\matx{\Psi^+}$ where $\lambda_0>\lambda_1$ by the local gauge transformations $U$ whereas the remaining link $ac$ can only be brought to the form $\lambda_0\matx{\Phi^+}+\lambda_1\matx{\Phi^-}$. The links $ab$ and $bc$ are in the desired form i.e. they correspond to $\Sigma$ with signature $(+,-,-,-)$, these are Bell diagonal states whose largest contribution in the Bell state mixture is the singlet. We have chosen this particular signature as it makes the parallel transporter $\Lambda$ unique provided $S$ is full rank and seems most natural (the spatial components of $S$ corresponding to $\sigma_x$, $\sigma_y$ and $\sigma_z$ are all treated equally). The singlet $\ket{\Psi^-}$ is also unique amongst the Bell states since it is the only anti-symmetric state whereas the other three Bell states are symmetric. We found that one cannot put every two qubit link in the form $\lambda_0\matx{\Psi^-}+\lambda_1\matx{\Psi^+}$ consistently with local gauge transformations. %We interpret this result as a reflection of the fact that there is no anti-symmetric state of more than two qubits.
For the GHZ case we find two of the three links can be brought to mixtures composed of the singlet while the remaining link must necessarily be a mixture of the symmetric Bell states. One can see a similar thing happens with the W state case.

Note that each of the two qubit density matrices have a maximal rank of two as a consequence of the Schmidt decomposition of the pure three qubit state. This fact gives one enhanced freedom in choosing local gauge transformations $U$ and allows one not only to symmetrize each two qubit state simultaneously but also put each link in Bell diagonal form maximizing the entanglement in each link. Rank two density matrices correspond to a $\Sigma$ with form
\begin{equation}
\Sigma=\begin{pmatrix}
         y & 0 & 0 & 0 \\
         0 & -x & 0 & 0 \\
         0 & 0 & -x & 0 \\
         0 & 0 & 0 & -y \\
       \end{pmatrix}.
\end{equation}
The Lorentz singular values are given by $y=\tfrac{1}{2}(\lambda_0+\lambda_1)$ and $x=\tfrac{1}{2}(\lambda_0-\lambda_1)$ where $\lambda_i$ are the eigenvalues (given in non-increasing order) of $\sqrt{\sqrt{\rho}(\sigma_y\otimes\sigma_y)\rho^*(\sigma_y\otimes\sigma_y)\sqrt{\rho}}$, a familiar operator used in calculation of the concurrence \cite{ref:Wootters98}. Now one can see where the extra gauge freedom comes in. $\Sigma$ is composed of two degenerate blocks, degenerate with regard to the Minkowski metric. In other words $\Sigma$ is the direct product $y \sigma_z \oplus -x \mathbb{I}$. This means $\Sigma$ is invariant under transformations of the form $u\Sigma u^T$ where $u=SO(1,1)\oplus SO(2)$ giving the extra freedom in choosing the last gauge transformation.

The interpretation of the result prompts the question whether one can put each of the three two qubit links in a mixture of symmetric Bell states simultaneously. A mixture of symmetric Bell states for the case of rank two Bell diagonal states corresponds to $\Sigma=\text{diag}\{y,y,x,-x\}$ ($\rho=\lambda_0\matx{\Psi^+}+\lambda_1\matx{\Phi^+}$). One can indeed find $U$ that take each link to a state of this form simultaneously. However, it should be noted that the parallel transporter assigned to each link $\Lambda=V\eta W^T\eta$ is no longer unique since $\Sigma$ no longer has the Minkowski signature. We can however choose the $\Lambda$ closest to the identity, in some sense the `most parallel' transformation between the qubits. Including this extra rule one finds the overall transformation associated to this new rule of parallel transport is the identity.

As a final comment, we note that our result has similarity to the Wigner angle or Wigner rotation \cite{ref:VanWyk84,ref:Aravind97}. The product of two pure Lorentz boosts in different spatial directions is not equal to another pure Lorentz boost, rather it is a pure boost multiplied by a rotation, the Wigner rotation. Aravind has shown that the Wigner rotation can be thought of as a holonomy in rapidity space (hyperbolic space) resulting from the area enclosed by the two pure boosts and the pure boost connecting the end points. The magnitude of this rotation is given by the area of a hyperbolic triangle, the length of whose sides are given by the rapidities of each of the boosts.

%Meaning of the result, symmetric and anti-symmetric states. Extra gauge freedom. Change of rule of parallel transport to get the identity. No anti-symmetric state of three qubits? Even stronger form for pure states of three qubits. Each link can be put into the form that maximizes its entanglement simultaneously. Form of each link in density matrix basis. What about states that do not have full rank $S$ and product states? What happens here? Similarity to the Wigner angle \cite{ref:VanWyk84,ref:Aravind97}.
%
%In the GHZ example can put links $ab$ and $bc$ into state $p_1\matx{\Psi^-}+p_2\matx{\Psi^+}$ (where $p_1 > p_2$) but last link $ac$ is in state $p_1\matx{\Phi^+}+p_2\matx{\Phi^-}$.
%
%In W state example can put links $ab$ and $bc$ into state $\ket{\Psi^-}$ but last link $ac$ is then $\ket{\Psi^+}$.

\subsection{Mixed states of three qubits}

Mixed states of three qubits show a wide range of possible holonomies. Generically, they have a full $SO^+(1,3)$ holonomy group structure however they may be hard to solve analytically. Here we give some simple examples.

\subsubsection{Untwisted states}

There is an untwisted state
\begin{equation}
\tfrac{1}{6}\left(\matx{\Psi^-_{ab}}\otimes \mathbb{I}_c + \matx{\Psi^-_{bc}}\otimes \mathbb{I}_a + \matx{\Psi^-_{ac}}\otimes \mathbb{I}_b\right).
\end{equation}
For this particular state each two qubit link is the Werner state
\begin{eqnarray}
&&\tfrac{1}{6}\mathbb{I}+\tfrac{1}{3}\matx{\Psi^-}=\\
&&\tfrac{1}{2}\matx{\Psi^-}+\tfrac{1}{6}\left(\matx{\Psi^+}+\matx{\Phi^+}+\matx{\Phi^-}\right),\nonumber
\end{eqnarray}
corresponding to the correlation matrix $S=\tfrac{1}{2}\times\text{diag}\{1,-\tfrac{1}{3},-\tfrac{1}{3},-\tfrac{1}{3}\}$.
Each of the links has no entanglement, although this state is right on the cusp of being entangled - adding a infinitesimal amount more of the anti-symmetric Bell state $\ket{\Psi^-}$ to the mixture would result in an entangled state. One also notes this state is an equal mixture of symmetric and anti-symmetric Bell states.

There is also the complementary untwisted state
\begin{equation}
\tfrac{1}{2}\left(\matx{GHZ}+\matx{W}\right)
\end{equation}
where the states are $\ket{GHZ}=\sqrt{\tfrac{1}{3}}\ket{000}+\sqrt{\tfrac{2}{3}}\ket{111}$ and $\ket{W}=\sqrt{\tfrac{1}{3}}(\ket{001}+\ket{010}+\ket{100})$. Each two qubit correlation matrix has the form $S=\tfrac{1}{2}\times\text{diag}\{1,\tfrac{1}{3},\tfrac{1}{3},\tfrac{1}{3}\}$ i.e. it is the same as the example above except each link is in the signature $(+,+,+,+)$, that is each link is an equal mixture of the three symmetric Bell states, $\matx{\Phi^+}+\matx{\Phi^-}+\matx{\Psi^+}$. This state is conjectured to be the state of two qubits with maximal dissonance, that is, maximal quantum correlations that are not entanglement \cite{ref:Modi10}.

\subsubsection{$SO(1,1)$ holonomy from rank 3 bipartite density matrices}

The three qubit state
\begin{equation}\label{eq:rank3}
p\matx{GHZ}+(1-p)\matx{\psi}
\end{equation}
shows a wide range of interesting behavior depending on the values of its variables. Depending on these values, the overall transformation around the loop can either be the identity, a $\pi$ rotation or a general element of $SO(1,1)$. Each link can exist in three possible regions in parameter space, two having entanglement in the two qubit link and one region being separable. The regions are separated by two critical points such that one of the singular values $s_i=0$ resulting in $\det S=0$. The critical points in this example mark the transition from the link being entangled to separable. The pure states are defined in eq.~(\ref{eq:rank3}) are $\ket{GHZ}=\tfrac{1}{\sqrt{2}}(\ket{000}+\ket{111})$ and $\ket{\psi}=x\ket{001}+y\ket{010}+z\ket{100}+w\ket{111}$. We choose the amplitudes of $\ket{\psi}$ to be real and positive.

We solve this example analytically by finding the form of the diagonalizing matrices $V$ and $W$ for each link directly without prior use of gauge transformations. One may then have to apply a Pauli rotation to get $\Sigma$ into the correct signature. The Pauli rotation applied depends on which of the regions the bipartite link belongs which in turn depends on values of the variables $p$, $x$, $y$, $z$ and $w$. We solve for the link $ab$, solutions for the other two links follow in an analogous way. The form of the correlation matrix is
\begin{widetext}
\begin{equation}
S(b,a)=\frac{1}{2}\begin{pmatrix}
                    w^2+x^2+y^2+z^2 & 0 & 0 & (1-p)(-w^2+x^2+y^2-z^2) \\
                    0 & 2(1-p)(wx+yz) & 0 & 0 \\
                    0 & 0 & -2(1-p)(wx-yz) & 0 \\
                    (1-p)(-w^2+x^2-y^2+z^2) & 0 & 0 & p+(1-p)(w^2+x^2-y^2-z^2) \\
                  \end{pmatrix}.
\end{equation}
\end{widetext}
One may diagonalize this correlation matrix with $V_b=T(B\otimes B^*)T^\dag$ and $W_a=T(A\otimes A^*)T^\dag$ where
\begin{eqnarray}
A=\begin{pmatrix}
    \left(\tfrac{\epsilon(x)y}{\epsilon(w)z}\right)^{1/4} & 0 \\
    0 & \left(\tfrac{\epsilon(w)z}{\epsilon(x)y}\right)^{1/4} \\
  \end{pmatrix},\nonumber\\
B=\begin{pmatrix}
    \left(\tfrac{\epsilon(x)z}{\epsilon(w)y}\right)^{1/4} & 0 \\
    0 & \left(\tfrac{\epsilon(w)y}{\epsilon(x)z}\right)^{1/4} \\
  \end{pmatrix}
\end{eqnarray}
and $\epsilon(q)=\sqrt{p+2(1-p)q^2}$. The resulting diagonal matrix, $\Sigma$, has singular values
\begin{eqnarray}
s_0&=&\tfrac{1}{2}\epsilon(w)\epsilon(x)+(1-p)yz,\nonumber\\
s_1&=&(1-p)(yz+wx),\nonumber\\
s_2&=&(1-p)(yz-wx),\nonumber\\
s_3&=&\tfrac{1}{2}\epsilon(w)\epsilon(x)-(1-p)yz.
\end{eqnarray}
However, depending on the values of the variables the signs of the singular values may not be correct. There are three regions, $I$, $II$ and $III$, corresponding to the different signs $s_2$ and $s_3$ may take. $s_0 \geq s_1 \geq 0$ for all values. Region $I$ corresponds to the range of parameters in which $wx>yz$, with the first critical point occurring at $wx=yz$ when $s_2=0$. This region is entangled with concurrence $\mathcal{C}_I=2(1-p)(wx-yz)=-2 s_2$ however the signature of $\Sigma$ is $(+,+,-,+)$. We can bring it to the correct signature $(+,-,-,-)$ by applying a Pauli $y$ rotation to either $A$ or $B$, this choice does not change the form of the parallel transporter $\Lambda$ we assign to the link. If we choose to put the rotation on qubit $a$ we replace $A\rightarrow i A\sigma_y$.

Region $II$ is separable although it still has dissonance and this occurs over the range $wx<yz<\tfrac{\epsilon(w)\epsilon(x)}{2(1-p)}$ corresponding to $\det S>0$. In this case the signature of $\Sigma$ is correct, $(+,+,+,+)$, and we can construct the parallel transporter from $V$ and $W$ without applying a Pauli rotation.

Region $III$ is also entangled and lies in the range $yz>\tfrac{\epsilon(w)\epsilon(x)}{2(1-p)}$. The concurrence in this region is given by $\mathcal{C}_{III}=2(1-p)yz-\epsilon(w)\epsilon(x)=-2 s_3$. The signature of this region is $(+,+,+,-)$ therefore one needs to apply a Pauli $z$ rotation to one of the qubits.

The parallel transporters one assigns to each link are thus determined by the region that each two qubit link resides in. There are $3^3=27$ possible combinations assigning three regions to three links and therefore $27$ possible overall transformations (or holonomies) around the three qubit loop. Many of these combinations turn out give the same overall transformation however.

We investigate the possible forms by first making some observations about $V$ and $W$. They consist of two blocks, one spanning the identity and $z$ component which belongs to the group $SO(1,1)$ and a second block equal to the identity spanning $x$ and $y$. If the coefficients in the link are such that it belongs to region $I$ we need to apply a Pauli $y$ rotation to get the signature correct. This rotation takes one of $V$ or $W$ to the block diagonal form $O(1,1)\oplus \sigma_z$ and thus the parallel transporter assigned to a link residing in region $I$ also has this form. Applying the same arguments to regions $II$ and $III$, we find
\begin{eqnarray}
\Lambda_I&=&O(1,1)\oplus \sigma_z,\nonumber\\
\Lambda_{II}&=&SO(1,1)\oplus \mathbb{I},\nonumber\\
\Lambda_{III}&=&SO(1,1)\oplus -\mathbb{I}.
\end{eqnarray}
We can therefore observe that any loop with an \emph{odd} number of links in the entangled region $I$ will have an overall transformation belonging to $O(1,1)\oplus \sigma_z$. An element from this group can only have eigenvalues $\{1,-1,1,-1\}$ giving $\xi(abc)=0$. That is, the overall transformation is a $\pi$ rotation.

Calculation for the case when all three links are separable (all three in region $II$) yields a trivial total transformation, the identity, corresponding to $\xi(abc)=1$. From this case one can infer that combinations of region $II$ and region $III$ result in a total transformation of either the identity ($\xi(abc)=1$) or a $\pi$ rotation ($\xi(abc)=0$), the former case occuring if there is an even number of region $III$ links and the latter if there are an odd number of region $III$ links.

The only remaining cases are those that feature an even number of region $I$ links. These cases have non-trivial holonomy. We first look at the cases where all links are entangled that is $\Lambda_I(a,c)\Lambda_I(c,b)\Lambda_{III}(b,a)$ and the two other permutations of region $III$. The eigenvalues of this transformation are
\begin{widetext}
\begin{eqnarray}
\Lambda_I(a,c)\Lambda_I(c,b)\Lambda_{III}(b,a)= \text{diag}\left \{\tfrac{y\epsilon(z)}{z\epsilon(y)},-1,-1,\tfrac{z\epsilon(y)}{y\epsilon(z)}\right \} \hspace{0.5cm}&\Rightarrow& \hspace{0.5cm} \xi(abc)=\tfrac{(y\epsilon(z)-z\epsilon(y))^2}{4 yz \epsilon(y)\epsilon(z)},\\
\Lambda_I(a,c)\Lambda_{III}(c,b)\Lambda_{I}(b,a)= \text{diag}\left\{\tfrac{y\epsilon(x)}{x\epsilon(y)},-1,-1,\tfrac{x\epsilon(y)}{y\epsilon(x)}
\right\}\hspace{0.5cm} &\Rightarrow& \hspace{0.5cm}\xi(abc)=\tfrac{(x\epsilon(y)-y\epsilon(x))^2}{4 xy \epsilon(x)\epsilon(y)},\\
\Lambda_{III}(a,c)\Lambda_I(c,b)\Lambda_{I}(b,a)= \text{diag}\left\{\tfrac{x\epsilon(z)}{z\epsilon(x)},-1,-1,\tfrac{z\epsilon(x)}{x\epsilon(z)}
\right\}\hspace{0.5cm} &\Rightarrow& \hspace{0.5cm}\xi(abc)=\tfrac{(x\epsilon(z)-z\epsilon(x))^2}{4 xz \epsilon(x)\epsilon(z)}.
\end{eqnarray}
\end{widetext}
The final case, that of one unentangled link, $\Lambda_I(a,c)\Lambda_I(c,b)\Lambda_{II}(b,a)$, and its permutations are identical to the case above, however the eigenvalues taking the value $-1$ become $1$.

In this example the holonomy is determined by the location of the crictical points in each link. These critical points mark the transition from entangled to separable. One can verify analytically that any two qubit density matrix with rank three or smaller will also have critical points determined by the entangled to separable transition. However, for two qubit density matrices of rank four this behaviour no longer holds as we illustrate in the next example.

The special role of the singlet in the holonomy is also highlighted again in this example. It is interesting and strange to note one can put all links simultaneously into states that do not have any singlet content. That is, one can make each $\Sigma$ a mixture of the three symmetric Bell states $\lambda_0\matx{\Psi^+}+\lambda_1\matx{\Phi^-}+\lambda_2\matx{\Phi^+}$ and the holonomy will be the identity no matter which of the regions each of the two qubit links resides. This is not possible if one requires any singlet content, the holonomy is always a $\pi$ rotation or a more general element of $SO(1,1)$. We believe this is due to the malleability, or size of the the symmetric state space; the ordering of the the three triplet states does not change the form of the parallel transporters. The singlet is much more rigid, its ordering will change the parallel transporters.

\subsubsection{$SO(1,1)$ holonomy from rank 4 bipartite density matrices}

An example of a state with $SO(1,1)$ holonomy with two qubit density matrices that are each full rank is the following state
\begin{equation}
p\matx{W}+(1-p)\matx{\overline{W}}.
\end{equation}
The state $\ket{W}$ is defined in eq.~(\ref{eq:wstate}) and $\ket{\overline{W}}=\sigma_x^{\otimes 3}\ket{W}$. In the following we take $w=0$ (choosing $w\neq 0$ results in a $SO^+(1,2)$ holonomy) and $p\in[0,1]$.

For these states we chose to solve for the parallel transporters directly without the prior use of gauge transformations. That is, we find the $V$ and $W$ for each two qubit link in the Lorentz SVD given by eq.~(\ref{eq:LSVD}). One finds the subspace of each $S$ spanned by $\sigma_x$ and $\sigma_y$ is already diagonal and the singular values associated to this subspace are degenerate. Only the subspace spanned by $\mathbb{I}$ and $\sigma_z$ needs to be diagonalized. This can be achieved by simple Lorentz boosts applied in the $z$ axis given by the group $SO(1,1)$. Explicitly each $V$ and $W$ has the form
\begin{equation}\label{eq:boost}
\begin{pmatrix}
  \cosh \varphi & 0 & 0 & \sinh \varphi \\
  0 & \pm 1 & 0 & 0 \\
  0 & 0 & \pm 1 & 0 \\
  \sinh \varphi & 0 & 0 & \cosh \varphi \\
\end{pmatrix}
\end{equation}
and therefore the parallel transporters and the overall transformation around the loop have the same form. That is, the holonomy is a simple pure boost along some spatial axis. Two of the singular values for each link are degenerate and take the same sign. The last spatial singular value associated to the boost however can change sign. It is negative when $\det S<0$ in which case the central diagonal block in eq.~(\ref{eq:boost}) becomes $-\mathbb{I}$ and $\mathbb{I}$ when $\det S>0$. This critical point occurs at $x^2\sqrt{p(1-p)}=\sqrt{p(1-p)(y^2-z^2)^2+y^2z^2}$ for the link $ab$ with analoguous relations for the other two links. There are therefore two regions seperated by the sign change of this singular value. The sign change in this example does not mark the transition between entangled and separable since the two qubit density matrices are rank 4 rather than rank 3. However, we can relate this point to (negative) concurrence; at the critical value the Bell state with the smallest probability in mixture represented by $\Sigma$ has probability $p_3=-2\mathcal{C}$ where $\mathcal{C}=\lambda_0-\lambda_1-\lambda_2-\lambda_3$ is negative at the critical point.

It is not hard to solve for this case. One finds the twist is given by hyperbolic cosine of the sum of the hyperbolic angles, $\varphi$,
\begin{equation}
\xi(abc)=\cosh^2\left(\frac{\varphi_{ac}+\varphi_{cb}+\varphi_{ba}}{2}\right)
\end{equation}
when there are an even number of links with $\det S<0$ and
\begin{equation}
\xi(abc)=\sinh^2\left(\frac{\varphi_{ac}+\varphi_{cb}+\varphi_{ba}}{2}\right)
\end{equation}
when there are an odd number of links with $\det S<0$. The rapidities (hyperbolic angles), $\varphi$, familiar variables from special relativity, are given by
\begin{equation}
\varphi=\frac{1}{2}\log\left(\frac{1+\beta}{1-\beta}\right)
\end{equation}
and each $\beta$ is
\begin{eqnarray}
\beta_{ba}=\frac{(1-2p)(y^2-z^2)}{y^2+z^2},\\
\beta_{cb}=\frac{(1-2p)(x^2-y^2)}{x^2+y^2},\\
\beta_{ac}=\frac{(1-2p)(z^2-x^2)}{z^2+x^2}.
\end{eqnarray}
Notice when $p\in\{0,\tfrac{1}{2},1\}$, $\xi=1$ i.e. one can untwist the bipartite correlations globally. One can also untwist the bipartite correlations for all $p$ if two or more of $x$, $y$ and $z$ are equal since the state before gauge transformations is more symmetrical.

\section{Conclusions}\label{sec:conclusions}

In this paper we have introduced a new measure quantifying the global degree of asymmetry of two qubit correlations in the spirit of a lattice gauge field theory. We call this measure twist and it is phrased as the familiar Wilson loop. Twist is defined as the trace of the total transformation or holonomy around a loop, a feature that cannot be `gauged' away by general local operations, the gauge group naturally emerging being the Lorentz group $SO^+(1,3)$ familiar from special relativity. This measure can be interpreted operationally in terms of an SLOCC protocol. We defined this measure as the trace of the total transformation which is a multipartite $SL(2,\mathbb{C})$ invariant of the state. However one may be able to get more information about the state by considering the eigenvalues of the total transformation which are also multipartite $SL(2,\mathbb{C})$ invariants. We point out that these invariants are not the polynomial invariants that have been typically studied in the literature. We have found that pure states of three qubits have a total transformation that is a $\pi$ rotation and provided some analytical examples of mixed three qubit states with more general transformations, elements of $SO(1,1)$, the group of Lorentz transformations in one dimension.

In this work we have been investigating the possible analogy between nonlocality in well known gauge field theory effects such as the Aharonov-Bohm effect and the nonlocality present in certain quantum states violating Bell inequalities. The aim being to phrase the nonlocality found in entangled states in terms of the familiar geometrical and topological ideas from gauge field theories. Our twist invariants are nonlocal, but not in the sense of states violating Bell inequalities. They are nonlocal in the sense that one may not attribute a twist to an individual link in the loop, since one may gauge it away to the trivial transformation. The twist is property of the entire loop of links and thus measures the global asymmetry of the two qubit correlations in that loop.

We have found that entanglement is not necessary to be able to define a unique parallel transporter to a link except in the case of rank 2 and lower two qubit density matrices. In the rank 3 case, entanglement does however define the crictical points in the link and these points affect the parallel transporter one assigns to a link. Generically however, to assign a unique parallel transporter to a link, one needs a weaker type of quantum correlation known in the literature as discord. Two qubit entanglement is determined by the magnitudes of the dilated correlations. One can phrase it as the sum of the singular values of the correlation matrix. Twist however, appears to characterize the asymmetry of the quantum correlations. In constructing twist one makes use of the additional information about the state encoded in the local operations that bring the correlation matrix to Bell diagonal form.

The constructions used in this paper are all based on two qubit correlations. It is not clear that these ideas generalize simply to $d$ level systems or three and higher qubit correlations. For a two qubit correlation matrix there is a simple standard form; one can bring any two qubit state to a Bell diagonal state with local operations. It is not clear that this standard form exists for two qu$d$its or three or more qubits. In the case of three qubits one has a $4\times 4 \times 4$ correlation tensor. One can imagine several possible standard forms for such an object.

We have several questions motivated by the similarity of this work to other ideas in modern theoretical physics. For example, in loop quantum gravity, the primitive variables are also phrased as Wilson loops, however parallel transporters are assigned from the relation between the discrete points of space rather than the relation between qubits \cite{ref:BaezBook}. One possible avenue of investigation could be to look for states with bipartite correlations that mimic simple curved spacetimes. For example can one find a $N$ qubit state that has the property such that the larger the loop around a central qubit becomes the total transformation becomes more trivial? This could be a toy example of a simple curved spacetime such as the spherically symmetric Schwartzchild metric.

One may also look for links between knots and quantum correlations. Witten has phrased Wilson loops in gauge field theories with a Chern-Simons action in $2+1$ dimensions as knot invariants \cite{ref:Witten89}. One wonders whether one can find states where a similar analogy can be made in our construction.

On a more concrete note, we hope that the ideas presented in this paper are useful in constructing and gaining an intuition for invariant correlation properties in $N$ qubit states.

\begin{acknowledgments}
We thank Johan {\AA}berg, Stephen Brierley, \v{C}aslav Brukner, Berge
Englert, Richard Jozsa, Noah Linden, Ognyan Oreshkov, Jiannis Pachos, Wonmin Son, Andreas Winter and especially Bill Wootters for helpful
comments and discussions. For financial support MSW acknowledges
EPSRC, QIP IRC www.qipirc.org (GR/S82176/01), the NRF and the MoE (Singapore) and an Erwin Schr\"{o}dinger JRF. ES and VV acknowledge the National Research Foundation and the Ministry of Education (Singapore). ME acknowledges support from the Swedish Research Council (VR).
\end{acknowledgments}

\bibliography{../../../references/masterbib}

\end{document}